\documentclass[journal]{vgtc}                     

\onlineid{1032}

\vgtccategory{Theoretical \& Empirical}

\vgtcpapertype{evaluation}

\title{Stop Misusing t-SNE and UMAP for Visual Analytics}

\author{%
  Hyeon Jeon, Jeongin Park, Sungbok Shin, and Jinwook Seo
}

\authorfooter{
  \item Hyeon Jeon, Jeongin Park and Jinwook Seo are with Seoul National University. E-mail: \{hj, jipark\}@hcil.snu.ac.kr, jseo@snu.ac.kr
  \item
  	Sungbok Shin is with Sogang University. E-mail: sbshin90@sogang.ac.kr
    \item Jinwook Seo is the corresponding author.
}

\abstract{%
Misuses of \tsne and \umap in visual analytics have become increasingly common.
For example, although \tsne and \umap projections often do not faithfully reflect the original distances between clusters, practitioners frequently use them to investigate inter-cluster relationships.
We investigate why this misuse occurs, and discuss methods to prevent it.
To that end, we first review 136 papers to verify the prevalence of the misuse.
We then interview researchers who have used dimensionality reduction (DR) to understand why such misuse occurs. 
Finally, we interview DR experts to examine why previous efforts failed to address the misuse. 
We find that the misuse of \tsne and \umap stems primarily from limited DR literacy among practitioners, and that existing attempts to address this issue---mostly based on academic papers---have been ineffective.
Based on these insights, we discuss potential future research directions to mitigate the misuse.

}


\teaser{
  \centering
  \includegraphics[width=\linewidth]{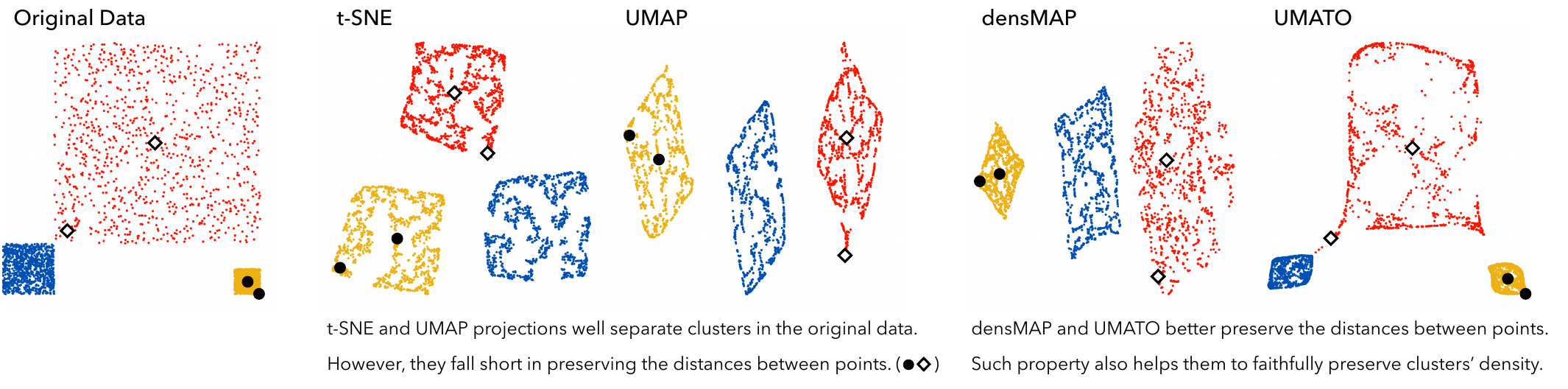}
  \caption{%
  	\textit{Comparison of \tsne, \umap, densMAP \cite{narayan21nature}, and UMATO \cite{jeon22vis} projections of a 2D dataset.} Although \tsne and \umap do not faithfully represent cluster density or inter-cluster distances, they are often misused to analyze such structures. Our research investigates why this misuse happens and explores strategies to mitigate these issues.
  }
  \label{fig:teaser}
}

\graphicspath{{figs/}{figures/}{pictures/}{images/}{./}} 

\usepackage{tabu}                      
\usepackage{booktabs}                  
\usepackage{lipsum}                    
\usepackage{mwe}                       

\usepackage{mathptmx}                  

\usepackage{amsmath}
\usepackage{amssymb}
\usepackage{enumitem}
\usepackage{listings}
\usepackage[dvipsnames, table]{xcolor}
\usepackage{makecell}
\usepackage{longfbox}
\usepackage{xspace}
\usepackage{graphicx}
\usepackage{comment}
\usepackage{kotex}

\usepackage{float}
\usepackage{newfloat}
\DeclareFloatingEnvironment[
  fileext=lol,
  listname={List of Listings},
  name=Listing
]{listing}

\renewcommand{\paragraph}[1]{
\vspace{3pt}
\noindent
\textbf{#1.}
}

\newcommand{\paragraphit}[1]{
\vspace{3pt}
\noindent
\textit{#1.}
}

\def\tsne{\textit{t}-SNE\xspace}
\def\umap{UMAP\xspace}

\definecolor{appleredlight}{RGB}{255, 105, 97}
\definecolor{appleorangelight}{RGB}{255, 179, 64}
\definecolor{appleyellowlight}{RGB}{255, 212, 38}
\definecolor{applegreenlight}{RGB}{48, 219, 91}
\definecolor{applemintlight}{RGB}{102, 212, 207}
\definecolor{appleteallight}{RGB}{93, 230, 255}
\definecolor{applecyanlight}{RGB}{112, 215, 255}
\definecolor{applebluelight}{RGB}{64, 156, 255}
\definecolor{appleindigolight}{RGB}{125, 122, 255}
\definecolor{applepurplelight}{RGB}{218, 143, 255}
\definecolor{applepinklight}{RGB}{255, 100, 130}
\definecolor{applebrownlight}{RGB}{181, 148, 105}

\definecolor{applerednormal}{RGB}{255, 69, 58}
\definecolor{appleorangenormal}{RGB}{255, 159, 10}
\definecolor{appleyellownormal}{RGB}{255, 214, 10}
\definecolor{applegreennormal}{RGB}{48, 209, 88}
\definecolor{applemintnormal}{RGB}{99, 230, 226}
\definecolor{appletealnormal}{RGB}{64, 200, 224}
\definecolor{applecyannormal}{RGB}{100, 210, 255}
\definecolor{applebluenormal}{RGB}{10, 132, 255}
\definecolor{appleindigonormal}{RGB}{94, 92, 230}
\definecolor{applepurplenormal}{RGB}{191, 90, 242}
\definecolor{applepinknormal}{RGB}{255, 55, 95}
\definecolor{applebrownnormal}{RGB}{172, 142, 104}

\newcommand{\redt}[1]{\textcolor{appleredlight}{#1}}

\newcommand{\redn}[1]{\textcolor{applerednormal}{#1}}
\newcommand{\bluen}[1]{\textcolor{applebluenormal}{#1}}

\def\oone{\textsc{O1}\xspace}

\def\otwo{\textsc{O2}\xspace}

\def\othree{\textsc{O3}\xspace}

\newcommand{\revise}[1]{#1}

\begin{document}

\definecolor{AnglePink}{RGB}{220,60,120} 


\lstdefinelanguage{json}{
  basicstyle=\ttfamily\small,
  showstringspaces=false,
  breaklines=true,
  breakatwhitespace=true,
  numbers=none,
  backgroundcolor=\color{white},
  frame=none,
  string=[s]{"}{"},
  stringstyle=\color{teal},
  keywords={true,false,null},
  keywordstyle=\color{blue}\bfseries,
  morecomment=[l][\color{gray}]{//},
  literate=
   *{0}{{{\color{Violet}0}}}{1}
    {1}{{{\color{Violet}1}}}{1}
    {2}{{{\color{Violet}2}}}{1}
    {3}{{{\color{Violet}3}}}{1}
    {4}{{{\color{Violet}4}}}{1}
    {5}{{{\color{Violet}5}}}{1}
    {6}{{{\color{Violet}6}}}{1}
    {7}{{{\color{Violet}7}}}{1}
    {8}{{{\color{Violet}8}}}{1}
    {9}{{{\color{Violet}9}}}{1}
    {.}{{{\color{Violet}.}}}{1}
    {-}{{{\color{Violet}-}}}{1}
    ,
  moredelim=**[is][\color{AnglePink}]{<}{>}
}

\lstdefinestyle{jsonstyle}{
  language=json,     
  tabsize=2,
  captionpos=b,
  escapeinside={(*@}{@*)}, 
  mathescape=false,
    columns=fullflexible,
  keepspaces=true,
  upquote=true
}

\renewcommand{\lstlistingname}{Code}


\firstsection{Introduction}

\maketitle


\label{sec:intro}

We take a close look at a widely known phenomenon threatening the reliability of visual analytics: \textit{the misuse of \tsne and \umap.} 
When practitioners refer to dimensionality reduction (DR) for visually analyzing high-dimensional data, \tsne and \umap are the most widely applied techniques \cite{espadoto21tvcg, jeon25chi, xuan22tvcg, hardvis23cgf, li23tvcg}.
However, these two are also commonly misused in practice \cite{chari23plos, distill2016how, coenen19fiar, cashman25arxiv}, \revise{i.e., \textit{used for analytical tasks that investigate the structure of high-dimensional data that the techniques do not aim to preserve.}}
For example, although \tsne and \umap do not accurately represent global structures like distances between points \cite{nonato19tvcg, jeon24tvcg, espadoto21tvcg, distill2016how}, they are often used to investigate the dissimilarity between data points or clusters \cite{chari23plos, distill2016how, cashman25arxiv} (\autoref{fig:teaser}).
Such misuse may introduce errors into visual analytics, which can propagate through interconnected visualizations and compromise their reliability.



We systematically investigate this misuse to understand how to address it. 
First, we verify the existence of the misuse by reviewing 136 visual analytics papers that utilize DR. 
Then, we conduct interviews with 12 researchers who frequently use DR for visual analytics (across data visualization, machine learning, and bioinformatics) to better understand the underlying causes of the misuse.  
As a final step, we interview eight DR researchers to understand why previous attempts have failed to fully resolve the underlying causes of this misuse. 



Our findings indicate that the misuse mainly occurs because practitioners have limited DR literacy.
Several participants in our first interview study consider \tsne and \umap to be ``immune to criticism,'' implying that both authors and reviewers lack sufficient knowledge of how to use DR appropriately. 
We also find that previous efforts to improve DR literacy have been ineffective.
Researchers across various domains produce papers that warn of the weaknesses of \tsne and \umap and that emphasize how to use DR properly (\autoref{sec:weaknesses}).
Unfortunately, such efforts rarely motivate practitioners to meaningfully engage with the issue and hence do not enhance DR literacy.






A plausible direction to mitigate misuse is to leverage automation to guide practitioners with limited DR literacy to use DR properly. 
However, the optimal level of automation for achieving these goals remains unknown.
Fully automating the selection of optimal DR techniques and hyperparameters may help practitioners who are entirely new to DR, but it can discourage them from acquiring DR literacy over time. 
At the opposite extreme, granting practitioners full agency over DR requires them to configure everything manually, making it both unrealistic and ineffective. 
Such a tradeoff highlights the need to identify the optimal mixed-initiative approach that leverages machine interventions to help practitioners avoid misuse while preserving their authority.

In summary, our research provides three key contributions:
\begin{itemize}
    \item We present a \textbf{literature review} and  \textbf{two interview studies} that investigate the misuse of \tsne and \umap.
    \item We verify the \textbf{prevalence of such misuse} and understand why previous approaches are ineffective in addressing the misuse. 
    \item We discuss \textbf{future research directions} to prevent practitioners from misusing DR and to enhance their DR literacy. 
\end{itemize}
We hope this research will spark discussions and encourage the appropriate use of not only DR but also other machine learning techniques, ultimately enhancing the reliability of visual analytics.

\section{Backgrounds and Related Work}

We discuss how the two DR techniques, \tsne and \umap, impact visual analytics research.
We then detail previous efforts in visualization, machine learning, and bioinformatics fields to address this problem.


\subsection{How \tsne and \umap Impact Visual Analytics}

We first explain DR, \tsne, and \umap.
Then, we discuss how the two techniques influence visual analytics. 

\paragraph{Dimensionality reduction}
DR, e.g., $t$-SNE, UMAP, PCA \cite{pearson01pmjs}, is an essential tool for visually analyzing high-dimensional data \cite{nonato19tvcg, endert12chi, fujiwara21tvcg, kuo22pvis, jeon25chi}. 
These techniques receive a high-dimensional dataset as input and produce a 2D representation that preserves important characteristics (e.g., local neighborhood structure or distances between clusters) of the original data. 
Using DR, any high-dimensional data can be visualized using a single scatterplot. Such effectiveness makes DR widely used for visual analytics across diverse domains \cite{cashman25arxiv}, including bioinformatics \cite{cheng23tvcg}, machine learning \cite{kahng18tvcg, hohman20chi}, and finance \cite{chen24tvcg}.

\paragraph{\tsne}
Since its first introduction in 2008 \cite{maaten08jmlr}, $t$-SNE has become one of the most widely used DR techniques.
It projects high-dimensional data by minimizing the divergence between two distributions: one representing pairwise similarities of the points in the original space and the other in the low-dimensional space \cite{lee11pcs, maaten08jmlr}. 

\paragraph{\umap} 
This technique is introduced in 2018 \cite{mcinnes2020arxiv} and has quickly gained popularity in diverse fields, including visual analytics.
UMAP captures the local structure of high-dimensional data by constructing a $k$-nearest neighbors ($k$NN) graph. 
It then optimizes a projection by minimizing the cross-entropy between the fuzzy topological representations of $k$NN graphs in the high and low-dimensional spaces.

\paragraph{Impact of \tsne and \umap in visual analytics}
These two techniques significantly influence the visualization and visual analytics fields by motivating numerous follow-up studies. 
They are frequently utilized in visual analytics systems \cite{xuan22tvcg, hardvis23cgf, li23tvcg}. These systems typically use projections to provide an overview of the data, allowing users to select subsets via interactions such as brushing and conduct follow-up analysis.
\revise{However, these techniques primarily preserve local structure and can therefore produce misleading representations of global structure, such as cluster density (\autoref{fig:teaser}). Nevertheless, they are frequently used in ways that risk such misinterpretation. Our work aims to verify this problem by understanding its underlying causes.}


\subsection{Previous Efforts to Address the Misuse}
\label{sec:weaknesses}

We identify literature from the visualization, machine learning, and bioinformatics domains that contributes to addressing the misuse of \tsne and \umap.



\paragraph{Quantitative experiments}
These works execute experiments to compare the suitability of the projections generated by diverse DR techniques on different analytic tasks. 
For example, Xia et al. \cite{jiazhi21tvcg} conduct a user study to test the effectiveness of  DR techniques in supporting cluster analysis. They reveal that $t$-SNE and UMAP are most effective for cluster identification tasks but are less effective for tasks such as density or distance comparison.  
Ventocilla and Reveiro \cite{bentocilla20ivs} investigate the alignment between human task accuracy in cluster analysis and clustering metrics, reaching a similar conclusion. 
Jeon et al. \cite{jeon24tvcg} and Lause et al. \cite{lause24biorxiv} show that $t$-SNE and UMAP work poorly for investigating global structures like cluster density or separability. 

\paragraph{Alternative DR techniques}
Diverse research proposes new DR techniques as alternatives to \tsne and \umap that mitigate their weaknesses in representing the original high-dimensional data.
Narayan et al. \cite{narayan21nature} verify that \tsne and \umap poorly represent cluster density and propose den-SNE and densMAP as alternatives. 
Trimap \cite{amid22arxiv}, PacMAP \cite{wang21jmlr}, and UMATO \cite{jeon22vis} improve \umap in terms of capturing the global structure (e.g., pairwise distances between data points) of the original data. Global \tsne \cite{zhou22nc} improves \tsne in the same direction. 
\autoref{fig:teaser} illustrates the superiority of alternative techniques against \tsne and \umap in preserving global structures.

\paragraph{Guidelines for proper use of \tsne and \umap}
These studies inform the limitations of \tsne and \umap to the public, guiding practitioners to use these techniques more appropriately. 
Wattenberg et al. \cite{distill2016how} provide guidance on appropriately using \tsne, and Coenen and Pearce \cite{coenen19fiar} offer similar insights for \umap. 
These works caution practitioners against relying on the global structure presented by \tsne and \umap and emphasize the substantial impact of hyperparameter selection on the faithfulness of resulting projections. 
Kobak et al. \cite{Kobak2021} show that initialization severely affects the faithfulness of the resulting projection and recommend using PCA for initializing \tsne and \umap. 
\revise{Kobak and Berens \cite{kobak2019art} recommend using a high learning rate and computing similarity over a diverse range to overcome the limitations of \tsne.}



\revise{
Despite such prior efforts to address misuse, both \tsne and \umap remain widely misused. 
We attribute this to practitioners' lack of motivation to cultivate DR literacy.
We propose solutions to address this problem, ranging from automating the process of finding optimal projections to actively facilitating discourse on the proper use of DR.
}

\section{Research Objectives}

We aim to address the misuse of \tsne and \umap.
We set \revise{three} research objectives to achieve this goal. 



\paragraph{(\oone) Verify the misuse of \tsne and \umap}
We want to find evidence that \tsne and \umap are widely used in visual analytics and are more frequently misused than alternative techniques. 
We thus want to provide a rationale for focusing on these two techniques.

\paragraph{(\otwo) Understand why practitioners misuse these techniques}
We aim to investigate the underlying cause of the misuse.
For example, we want to investigate the \revise{rationales} behind the misuse of \tsne and \umap.
This investigation grounds our suggestions for future directions to mitigate the misuse.

\paragraph{(\othree) Understand why previous efforts fall short in addressing the misuse}
We investigate why existing efforts have failed to prevent the persistent misuse of t-SNE and UMAP.
As with \otwo, this investigation supports our suggestion of a new strategy to overcome the limitations of previous approaches. 



\vspace{3pt}

The remaining parts of this paper are dedicated to achieving these objectives. First, we conduct a literature review (\autoref{sec:litreview}) on visual analytics papers using DR to investigate the extent to which \tsne and \umap are misused (\oone).
We deepen our investigation into misuse through an interview study with practitioners who regularly use DR (\autoref{sec:interstudy}), observing the underlying causes of the misuse (\otwo). 
We then interview expert researchers who study DR (\autoref{sec:interviewexperts}) to obtain insights on why previous approaches are not effective in mitigating the misuse (\othree). 
Based on the findings, we propose future directions to address the misuse of \tsne and \umap in visual analytics (\autoref{sec:checklist}).

\section{Literature Review}

\label{sec:litreview}

We execute a literature review to confirm that \tsne and \umap are commonly misused in visual analytics (\oone).
We confirm that (\textbf{H1}) \tsne and \umap dominate the use of DR in visual analytics, providing a rationale for our focus on these two techniques. 
We also hypothesize and verify two types of misuse: (\textbf{H2}) the use of \tsne and \umap for unsuitable tasks, and (\textbf{H3}) the lack of appropriate justification. 




\subsection{Protocol}

\label{sec:protocol}

The review consists of four steps: paper search, categorization, task suitability review, and quantitative analysis. 

\paragraph{Paper search}
We search for \textit{papers that propose a visual analytics approach, framework, or systems that incorporate the \revise{existing DR techniques}}.
We query papers that satisfy two conditions: (1) the term \texttt{``visual analytics''} or \texttt{``visual analysis''} appears in the title or abstract, and (2) one or more of the terms among \texttt{``dimensionality reduction''}, \texttt{``dimension reduction''}, \texttt{``multidimensional projection''}, and \texttt{``multidimensional scaling''} are mentioned in the full text. 
We use IEEE Xplore, Wiley online library, and Sagepub to search papers published in major data visualization journals and conferences (e.g., IEEE VIS, TVCG, CG\&A, PacificVis, EuroVis, CGF, IVIS).
We filter out papers published before 2008, the year \tsne was announced.
Then, we inspect each paper and remove those that do not fall within our search scope (e.g., papers that propose novel DR techniques or pipelines \cite{fujiwara20tvcg, kwon17tvcg} or focus on other kinds of data, e.g., networks \cite{shi22tvcg, col18tvcg}).
Two authors engage in this process; we exclude a paper only when both agree to minimize false negatives. The eligibility of the papers is determined by the title, abstract, introduction, and methods sections.
The list of identified papers is in Appendix A.


\paragraph{Categorization}
We categorize the identified papers according to the following three criteria:

\paragraphit{DR techniques}
We mark a paper as using a certain DR technique only when the paper explicitly mentions that they use the technique to visualize high-dimensional data. We exclude cases in which DR is used solely for data preprocessing or compression \cite{jeon25chi, fujiwara20tvcg}, as our goal is to determine whether DR techniques, including \tsne and \umap, are used for unsuitable visual analytics tasks or not (H2).

\paragraphit{Target analytic tasks}
We categorize a paper as performing a specific analytic task when it states that the task is an intended target of the analysis.
Meanwhile, several papers do not explicitly specify their target analytic tasks. In such cases, we examine user studies, case studies, and use cases to infer the target tasks. For instance, if a case study includes analyses that ``lasso'' a cluster for detailed investigation, we classify the paper as performing a cluster identification.

\paragraphit{\revise{Rationales}}
We examine the system design of each paper to identify the stated rationale for selecting particular DR techniques. A paper is marked to have a specific \revise{rationale} when the rationale is explicitly stated. If no \revise{rationale} is provided, we label the paper as ``No reason''.

\vspace{4pt}
\noindent
We select the first two criteria to investigate the extent to which DR techniques are misused for tasks that are not suitable for them (H2).
We select \revise{rationales} as an additional criterion to check whether they justify the use of \tsne, \umap, and other DR techniques in an inappropriate way (H3). 
We follow the thematic coding process for the categorization.
Two coders first independently categorize the papers, where the initial agreement measured by Cohen's $\kappa$ is 0.62. 
They merge and revise their categorizations through three iterative discussions (\autoref{sec:categorization}).

\paragraph{Task suitability review}
We assess the suitability of major DR techniques for different analytic tasks 
by examining prior work that evaluates the weaknesses and strengths of different DR techniques \cite{jiazhi21tvcg, jeon22vis, jeon24tvcg, narayan21nature} (\autoref{sec:weaknesses}). 
\revise{Following our definition of misuse}, we consider a DR technique suitable for an analytic task if it preserves the structural characteristics relevant to that task. When suitable DR techniques are used, the corresponding analytic tasks can be conducted more reliably.
Two coders conduct the investigation through the following procedure.
At first, the coders review the papers independently and record the task suitability of the DR techniques by quoting the texts from the papers. We only consider claims supported by evidence such as computational benchmarks, user studies, or case studies. The coders then unified their findings to derive a final conclusion regarding the task suitability of the DR techniques. If conflicts arise among papers, we prioritize claims supported by stronger evidence (e.g., we place greater weight on findings from user studies than those from case studies) and by a larger number of papers.
We depict the results in \autoref{sec:suitability}.

\paragraph{Quantitative analysis}
We quantitatively analyze the categorized papers to verify H1 and H2. Detailed results are presented in \autoref{sec:quantianal}.

\subsection{Paper Search and Categorization Results}

\label{sec:categorization}

We retrieve 312 papers, and after screening and filtering, we retain 136 papers. 
The categories we identify from them are described below.

\subsubsection{DR Techniques}

We identify 18 DR techniques in total. 
Among them, we find four commonly used techniques (\tsne, \umap, PCA, and MDS) for which each is used more than 20 times. 
The other 14 techniques are used fewer than five times each, and we categorize them all under ``others.''

\begin{figure*}[t]
    \centering
    \includegraphics[width=\linewidth]{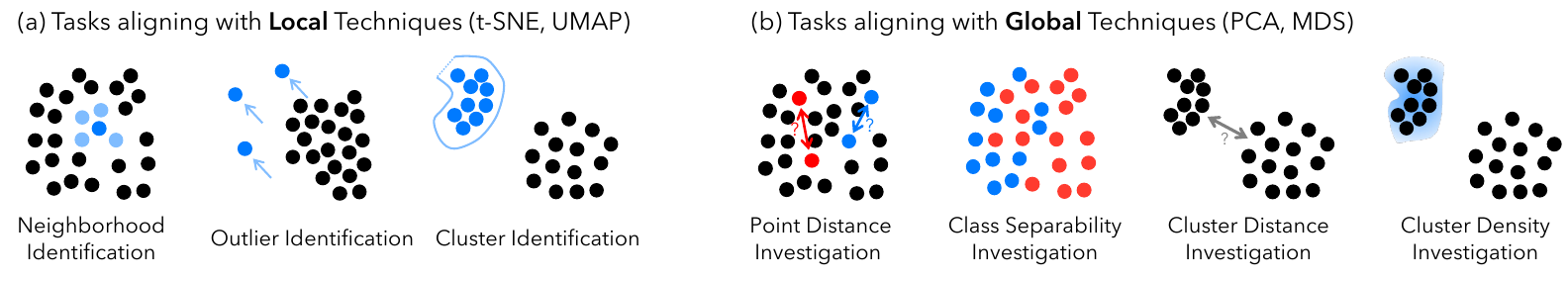}
    \vspace{-6mm}
    \caption{\textit{Illustrations of the analytic tasks using DR and their alignment to local and global DR techniques}. Our literature review identifies seven types of analytic tasks. \tsne and \umap are suitable for neighborhood, outlier, and cluster identification tasks, but are inappropriate for other tasks. }
    \label{fig:tasks}
\end{figure*}

\subsubsection{Analytic Tasks}

\label{sec:analtasks}

Our review reveals that analytic tasks using DR can be divided into seven categories (\autoref{fig:tasks}). 
The first three are \textit{identification} tasks: tasks that find or extract specific data points or their set. 
The other four are \textit{investigation} tasks: the tasks that measure or quantify characteristics of the data. Detailed descriptions of each task are as follows.


\paragraph{Neighborhood identification}
This task aims to find data points similar to a target point based on the proximity within the projection. 
Since this task supports many other analytic tasks, such as cluster identification, preserving local neighborhood structure is often considered the most important criterion when evaluating DR projections \cite{nonato19tvcg, jeon22vis, moor20icml}.

\paragraph{Outlier identification}
This task is about identifying outliers within projections. 
Analysts often count the number of outliers in the data \cite{etemadpour15tvcg} or determine whether a point is a cluster member or an outlier \cite{jiazhi21tvcg}. 
The task is typically used to examine the quality of class labels by identifying points with high uncertainty about their membership \cite{hong23pvis}.

\paragraph{Cluster identification}
This task involves identifying clusters within DR projections. 
Analysts typically count the number of clusters \cite{etemadpour15tvcg} or label clusters interactively using selection tools such as lasso or box-shaped brushes \cite{jiazhi21tvcg, meng24tvcg}. 
This task often includes investigating subclusters within existing clusters \cite{etemadpour15tvcg}. 
Visual analytics systems often provide auxiliary visualizations to show detailed information about the identified clusters \cite{chatzimparmpas20tvcg, li23tvcg}.

\begin{table}[t]
    \centering

    \caption{\textit{The coverage of tasks (rows) by references (columns).} We refer to each work using the last name of the first author. (Xia: Xia et al. \cite{jiazhi21tvcg}, Ete.: Etemadpour et al. \cite{etemadpour15tvcg}, Bre.: Brehmer et al. \cite{brehmer14beliv}, Non.: Nonato and Aupetit \cite{nonato19tvcg}, Sed.: Sedlmair et al. \cite{sedlmair12report}, Cas. : Cashman et al. \cite{cashman25arxiv})}
    \vspace{-1.2mm}
    \scalebox{0.85}{
    \begin{tabular}{r|cccccc}
    \toprule
    \textbf{Task} & \textbf{Xia} & \textbf{Ete.}  & \textbf{Bre.} & \textbf{Non.}  & \textbf{Sed.} & \textbf{Cas.}\\
    \midrule
         Neighborhood Identification &  & $\bullet$ & & $\bullet$ & & $\bullet$\\
         Outlier Identification &  $\bullet$  & $\bullet$ & & $\bullet$ & & $\bullet$\\
         Cluster Identification &  $\bullet$& $\bullet$ & $\bullet$ & $\bullet$ & $\bullet$ & $\bullet$\\
         Point Distance Investigation &  & & & $\bullet$ & & $\bullet$\\
         Class Separability Investigation & & & $\bullet$ & $\bullet$ & $\bullet$ & $\bullet$ \\
         Cluster Distance Investigation & $\bullet$ & $\bullet$ & $\bullet$ & $\bullet$ & $\bullet$ & $\bullet$\\
         Cluster Density Investigation&  $\bullet$ & $\bullet$& & $\bullet$\\
    \bottomrule
    \end{tabular}
    }
    \label{tab:tasks}
\end{table}

\paragraph{Point distance investigation}
Similar to the cluster distance investigation task, this task investigates the distance between data points as a proxy for their high-dimensional dissimilarity. 
It can be interpreted as a ``continuous'' version of the neighborhood identification task. 

\paragraph{Class separability investigation}
This task investigates how distinctly different classes are separated or distinct in the projections, where the classes are color-coded. 
It involves analyzing both the degree to which classes are mixed with each other and the relative distances among classes within the projection.
The task is commonly performed when DR techniques are employed to explain the behavior of a supervised machine learning model, particularly to illustrate how the model distinguishes between different classes \cite{ploshchik23pvis}.

\paragraph{Cluster distance investigation}
This task uses the distance between well-separated clusters as a proxy for their similarity in the original high-dimensional space.
The clusters can be explicitly labeled (i.e., color-coded) or implicitly represented by data distribution \cite{sedlmair12report}. 

\paragraph{Cluster density investigation}
This task identifies and compares the density of clusters using cluster density as a proxy for the variability of data points within each cluster \cite{narayan21nature}.


\vspace{4pt}
\noindent
\textit{Task coverage validation.}
To validate the comprehensiveness of our categorization, we examine whether the analytic tasks in our list are included in prior task taxonomies within the visualization field. 
We review Etemadpour et al. \cite{etemadpour15tvcg}, Xia et al. \cite{jiazhi21tvcg}, Brehmer et al. \cite{brehmer14beliv}, Nonato and Aupetit \cite{nonato19tvcg}, Sedlmair et al. \cite{sedlmair12report}, and Cashman et al. \cite{cashman25arxiv}. 
We find that all tasks are covered by at least two previous studies (\autoref{tab:tasks}), confirming that our categorization covers all important tasks in DR.

\renewcommand{\arraystretch}{1.1}
\begin{table*}[t]
    \centering
    
    \caption{\textit{The definition of the \revise{rationales} that we identify from the literature review (\autoref{sec:docreasonings})} 
    Except for extensibility and simplicity, all \revise{rationales} are leveraged to justify the use of \tsne and \umap. }
    \vspace{-1.2mm}
    \scalebox{0.91}{
    \begin{tabular}{ll}
    \toprule 
    \textbf{\revise{Rationale}} & \textbf{Definition} \\
    \midrule
     \textbf{Faithfulness} & The degree to which DR techniques accurately represent the original structure of the high-dimensional data without distortions. \\
        \textbf{Popularity} &  The degree to which DR techniques are widely known and used by practitioners in the visual analytics field. \\
        \textbf{Scalability} & Computational efficiency in executing DR techniques. \\
        \textbf{Interpretability} & The degree to which DR techniques yield visually distinct, analyzable clusters, enabling clear explanation of the data. \\
        \textbf{Stability} & The degree to which DR techniques produce projections that are stable against hyperparameter change or the stochastic nature of DR. \\ 
        \textbf{Extensibility} &  The degree to which DR techniques can be adapted or expanded to accommodate diverse data conditions or input formats. \\
        \textbf{Simplicity} & The degree to which practitioners can readily understand and apply DR techniques. \\
        \bottomrule 
    \end{tabular}
    }
    \label{tab:defreasonings}
\end{table*}
\renewcommand{\arraystretch}{1}

\subsubsection{\revise{Rationales}}
\label{sec:docreasonings}

We identify seven large categories of \revise{rationales} used to justify the selection of DR techniques. 
We define each \revise{rationale} in \autoref{tab:defreasonings}.
It is worth noting that a substantial number of papers (44\%; \autoref{fig:reasonings}) do not mention specific \revise{rationale}, i.e., marked as ``No reason''.

\paragraph{Faithfulness} 
Researchers justify the use of DR techniques based on their faithfulness, i.e., their ability to accurately represent the structure of high-dimensional data without distortion.
This \revise{rationale} mostly relies on references to benchmark studies that evaluate DR techniques~\cite{espadoto21tvcg, jiazhi21tvcg}. 
One notable finding is that researchers often cite the original UMAP paper \cite{mcinnes2020arxiv} to support claims about its capability to preserve global structure, which is not always correct \cite{coenen19fiar, jeon24tvcg2, jeon22vis, jiazhi21tvcg, narayan21nature}. We also find several papers claiming the faithfulness of UMAP without references. 

\paragraph{Popularity} 
Researchers justify the use of DR techniques based on their popularity, which indicates the degree to which the techniques are widely acquainted and used by practitioners in the visual analytics field.
For example, papers mention employing \tsne because it is a ``default option'' in visualizing high-dimensional data or is ``widely recognized'' by the research community. 
These papers also highlight specific research communities, such as biology, computer vision, and document clustering, where these techniques are commonly used \cite{cashman25arxiv}. 

\paragraph{Scalability}
The use of DR techniques is also justified by their
computational efficiency, which enhances the responsiveness of visual analytics systems. 
For example, papers state that efficient GPU implementations \cite{nolet22biorxiv, pezzotti20tvcg} facilitate the effectiveness of DR techniques.

\paragraph{Interpretability}
Researchers use DR techniques because they enable a clear explanation of the data with projections that contain visually distinct and, thus, analyzable clusters.
This finding aligns with the work by Morariu et al. \cite{morariu23tvcg} and Doh et al. \cite{doh25arxiv} that identify clear cluster separation as a key factor influencing the preference for DR projections.

\paragraph{Stability}
Researchers justify the selection of DR techniques by highlighting their stability (i.e., the degree to which DR techniques produce projections that are stable against hyperparameter change or the stochastic nature of DR) as a means to improve the reproducibility of their research. 
For example, one paper argues that \tsne is stable due to its non-convex optimization \cite{arora18clt}. 

\paragraph{Extensibility}
We find that researchers justify the use of DR techniques based on their extensibility or their ability to adapt or expand to accommodate diverse data conditions or input formats.
For example, some papers use DR techniques because they are parametric, i.e., support new data points to be dynamically projected after initial projection \cite{sainburg21nc}, particularly for analyzing streaming, online data. 

\paragraph{Simplicity}
A few papers mention selecting DR techniques that are simple and easy for practitioners to understand and apply. 
Among the four major techniques, only PCA has been justified by this \revise{rationale}.



\subsection{Suitability of DR Techniques to Analytic Tasks}

\label{sec:suitability}

We assess the suitability of four major DR techniques (\tsne, \umap, PCA, and MDS) to the analytic tasks identified in \autoref{sec:analtasks}. 
This is done by revisiting previous studies that evaluate DR techniques and analyze the alignment between the DR techniques and analytic tasks (\autoref{sec:weaknesses}) \revise{(e.g., \cite{espadoto21tvcg, nonato19tvcg, jeon22vis, moor20icml, zhou22nc, jiazhi21tvcg})}.
This analysis helps examine whether researchers are applying DR to tasks that are suitable (H1). 

Here, we clarify that the suitability of DR projections cannot be assessed solely based on the selection of DR techniques.
For example, hyperparameters such as perplexity in $t$-SNE can significantly affect how well a projection aligns with different analytic tasks.
Still, we adopt a coarse-grained approach that assesses suitability primarily by selecting DR techniques for the following reasons.
First, we identify that nearly all studies we investigated did not report their hyperparameter settings or other configurations, like seed selection. 
Second, the choice of DR technique remains the most influential factor affecting the task suitability of DR projections \cite{espadoto21tvcg, nonato19tvcg}.
Exploring task suitability by considering a broader range of factors would be an interesting avenue for future work. We discuss this direction in \autoref{sec:discussuse}.

\subsubsection{Tasks Suitable for \tsne and \umap}
\tsne and \umap focus on preserving local neighborhoods by placing nearby points close together while separating non-neighbors. They are thus commonly referred to as local techniques. Several studies demonstrate that they show state-of-the-art performance in preserving local structures, both empirically \cite{jeon22vis, espadoto21tvcg, moor20icml} and theoretically \cite{lee11pcs}. 
We categorize the following tasks that are suitable for \tsne and \umap.

\paragraph{Neighborhood identification task is more suitable for \tsne and \umap}
As aforementioned, \tsne and \umap directly aim to preserve local neighborhood structure. This makes them better suited for neighborhood identification tasks than alternative techniques by design \cite{venna10jmlr, amid22arxiv, jeon24tvcg, zhou22nc}, which also have been empirically validated \cite{jeon22vis, moor20icml}.

\paragraph{Outlier identification task is more suitable for \tsne and \umap}
Since projections generated by local techniques clearly distinguish neighboring and non-neighboring points, they can effectively separate outliers from clusters. 
Xia et al. \cite{jiazhi21tvcg} empirically show that \tsne and \umap are the most effective DR techniques for outlier identification, outperforming alternatives such as PCA.

\paragraph{Cluster identification task is more suitable for \tsne and \umap}
As \tsne and \umap locate neighboring points close and non-neighboring points far away \cite{jeon22vis, moor20icml, venna10jmlr}, they clearly represent individual high-dimensional clusters as 2D clusters, thus suitable for the cluster identification task. 
Xia et al. \cite{jiazhi21tvcg} show that the participants perform best when identifying clusters with \tsne and \umap.

\subsubsection{Tasks Suitable for PCA and MDS}
PCA and MDS are DR techniques that preserve global pairwise distances between data points more effectively than local techniques \cite{nonato19tvcg, jeon22vis, jiazhi21tvcg, sorzano14arxiv, van09jmlr}. They are usually referred to as global techniques. The following tasks are suitable for these DR techniques.

\paragraph{Point distance investigation task is more suitable for PCA and MDS}
These techniques are designed to directly preserve the pairwise distance structures more effectively compared to local techniques. They are thus more suitable than \tsne and \umap in investigating distances between data points (\autoref{fig:teaser} dots and diamonds). 
Several studies \cite{jeon22vis, moor20icml, amid22arxiv} propose techniques that improve UMAP in preserving global pairwise distances between points, such as UMATO \cite{jeon22vis} and TriMap \cite{amid22arxiv}.

\paragraph{Class separability investigation task is more suitable for PCA and MDS}
The superiority of PCA and MDS in preserving distances between data points makes them more precisely exhibit the separability between class labels \cite{distill2016how, jeon24tvcg, xia22tvcg}.
Local techniques like \tsne and \umap can well depict the extent to which classes are ``mixed'' \cite{jeon24tvcg}. However, they are widely reported to exaggerate the distance between classes \cite{bernard21cng, atzberger25tvcg, benato24cvicg, jeon24tvcg}. Wattenberg et al. \cite{distill2016how} show that hyperparameter choices can significantly distort class distances in \tsne.

\paragraph{Cluster distance investigation task is more suitable for PCA and MDS}
PCA and MDS better preserve pairwise distances between points within each cluster compared to alternatives, making the inter-cluster distances in the resulting projections meaningful. 
These techniques are thus suitable for tasks involving cluster distance analysis.
Many computational benchmarks have validated the superiority of PCA and MDS in supporting the cluster distance investigation task \cite{jeon22vis, moor20icml, wang21jmlr, bentocilla20ivs, jeon24tvcg}.
This implies the appropriateness of these techniques for supporting cluster distance investigation.  
Xia et al. \cite{jiazhi21tvcg} empirically show that global techniques like PCA enable users to perform this task more accurately than local techniques. 
In contrast, Wattenberg et al. \cite{distill2016how} and Coenen et al. \cite{coenen19fiar} also inform that the distance between clusters lacks meaning in \tsne and \umap projections, respectively.

\paragraph{Cluster density investigation task is more suitable for PCA and MDS}
These techniques depict the similarity between data points as low-dimensional proximity and thus can more sensitively depict the differences in cluster densities. 
In contrast, local techniques like \tsne and \umap poorly reflect their true similarity in high-dimensional space \cite{jeon22vis, amid22arxiv, narayan21nature} as they only focus on neighboring points. 
As a result, \tsne and \umap projections poorly represent cluster density (\autoref{fig:teaser}), which motivates the development of improved techniques such as den-SNE and densMAP \cite{narayan21nature}. 
The superiority of global techniques in supporting the density investigation task is also validated by Xia et al. \cite{jiazhi21tvcg} through user studies and Jeon et al. \cite{jeon24tvcg} via case studies.

\begin{figure}
    \centering
    \includegraphics[width=\linewidth]{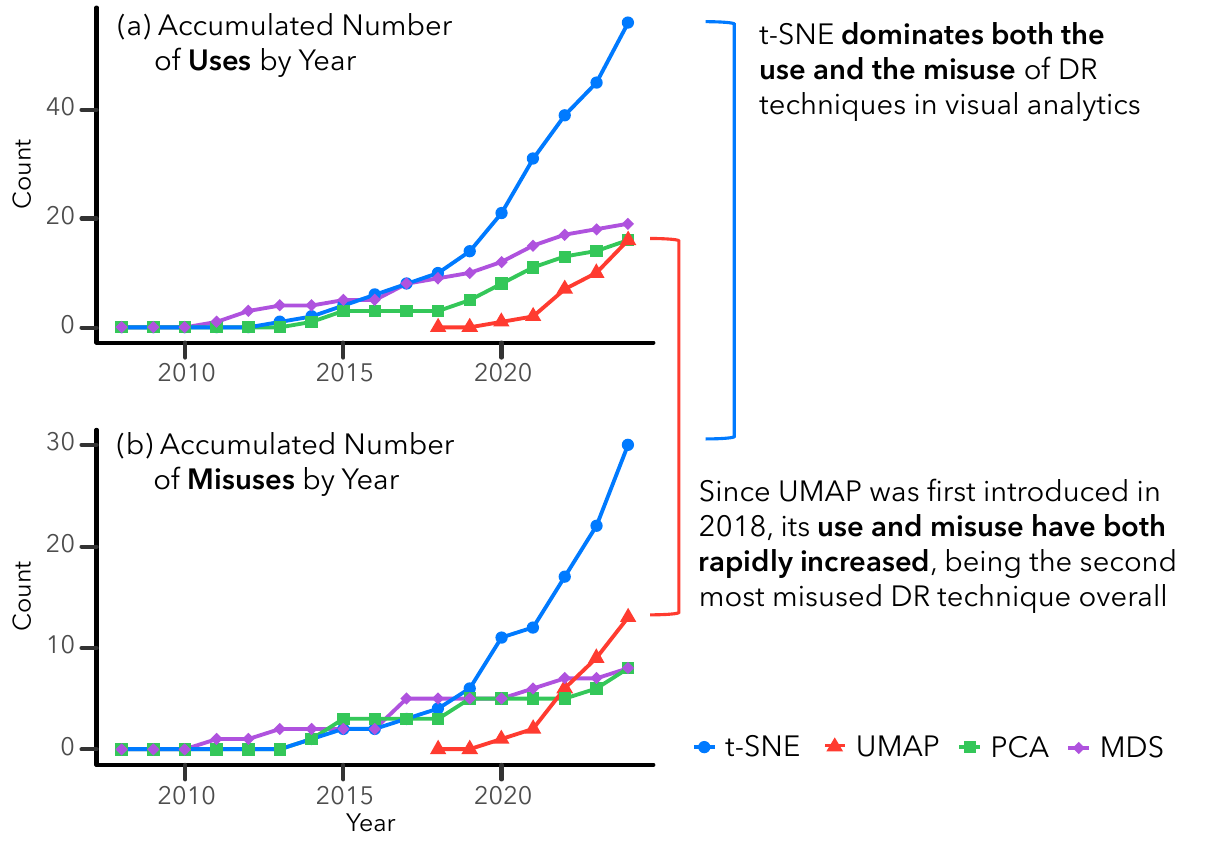}
    \vspace{-6mm}
    \caption{\textit{The trend of the accumulated number of papers that use (a) or misuse (b) of four major DR techniques.} We collect papers published from 2008, the year \tsne is introduced. UMAP's data also starts from the year it is released (2018). \revise{As these data are drawn from the set of papers we surveyed, the actual prevalence of DR use and misuse in practice may be higher than the reported numbers suggest.}}
    \label{fig:trend}
\end{figure}




\subsubsection{Validity of the Suitability Analysis} 
One notable finding in our suitability analysis is that \tsne and \umap perform better on all ``identification'' tasks while PCA and MDS excel at ``investigation'' tasks. 
This result aligns with the fundamental differences in how these methods interpret distances. \tsne and \umap prioritize preserving local neighborhoods by treating similarity as a binary function (neighbors or non-neighbors), making them well-suited to tasks that require identifying distinct clusters or groups. In contrast, PCA and MDS interpret distances as continuous values, enabling more accurate interpretation of relative distances between points. This finding supports the validity of our task categorization in capturing the alignment between DR techniques and the tasks that are suitable for.

\subsection{Findings}

\label{sec:quantianal}

We present the findings from the analysis of the identified papers (\autoref{sec:categorization}).
We reveal that \tsne and UMAP are the most commonly adopted DR techniques (H1). 
Yet, researchers often use them across any tasks, making them simultaneously the most commonly misused DR techniques (H2).
We also observe that many papers leverage \tsne and \umap without justifications or with improper rationale (H3).



\paragraph{(Finding 1) \tsne and \umap dominate the use of DR}
While the number of papers using DR has increased over the years, this growth is largely driven by \tsne and \umap (\autoref{fig:trend}; H1).
\tsne appears in more than half of the identified papers (75 out of 136), more than twice as often as the runner-up.
\umap is used in 31 papers. However, UMAP's adoption is increasing at a much steeper rate than PCA and MDS, enabling it to achieve parity with these established techniques in only six years.
These findings highlight that misusing \tsne and \umap can have a more harmful impact than misusing other techniques.


\begin{figure}[t]
    \centering
    \includegraphics[width=\linewidth]{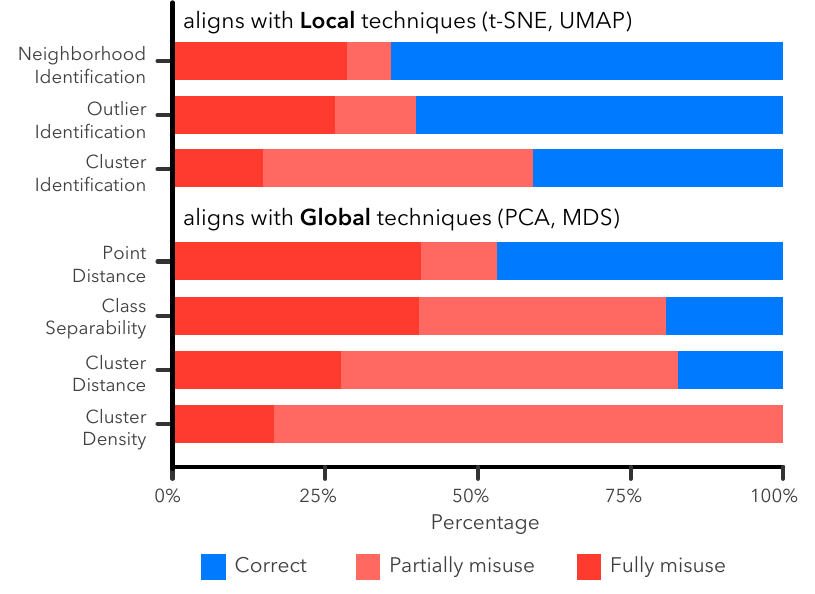}
    \vspace{-6mm}
    \caption{\textit{The ratio of appropriate use and misuse of DR techniques by analytic tasks.} DR is properly used for tasks that align with local techniques (top 3) but not for those that align with global techniques (bottom 4). This result indicates that \tsne and \umap are used even for unsuitable tasks. }
    \label{fig:error-tasks}
\end{figure}

\begin{figure}[t]
    \centering
    \includegraphics[width=\linewidth]{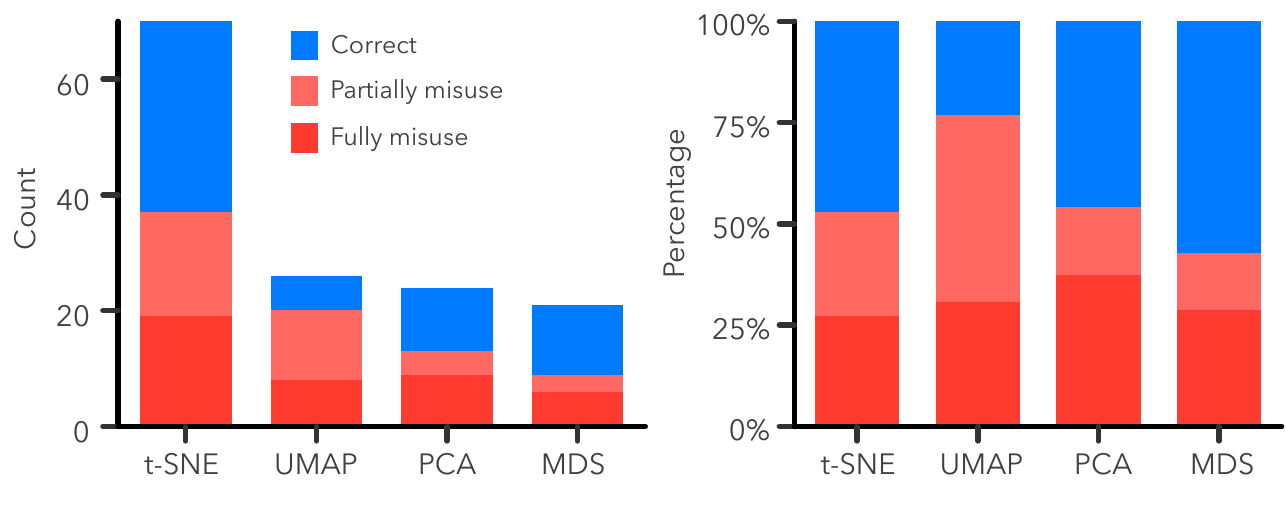}
    \vspace{-6mm}
    \caption{\textit{The number of appropriate uses and misuses of DR by techniques (left) and their ratio (right).} The analysis reveals that \tsne and \umap dominate the misuse of DR.}
    \label{fig:misuse_summary}
\end{figure}

\begin{figure*}
    \centering
    \includegraphics[width=\textwidth]{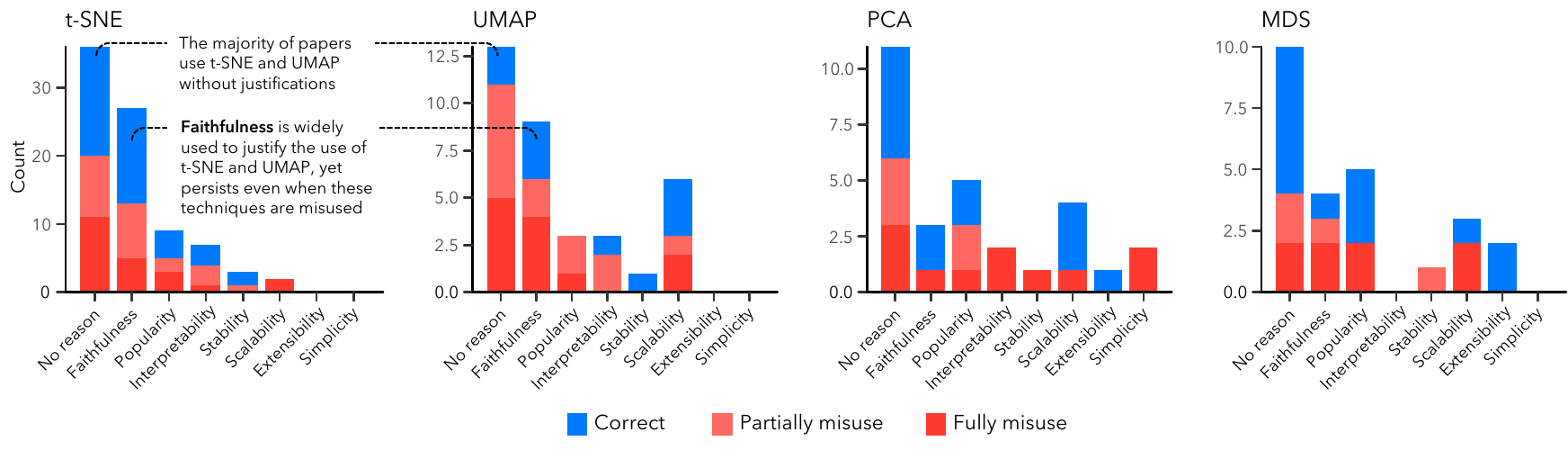}
    \vspace{-6mm}
    \caption{\textit{The number of appropriate uses and misuses of DR techniques by \revise{rationales}.} The \revise{rationales} (x-axis) are sorted in descending order based on the number they are referenced to justify the use of \tsne (leftmost chart). 
    }
    \label{fig:reasonings}
\end{figure*}

\paragraph{(Finding 2) \tsne and \umap are used for \textit{any} tasks}
We identify that \tsne and UMAP are commonly used for both suitable and unsuitable tasks, confirming H2. 
To do so, we determine whether a paper properly uses DR techniques for suitable tasks. Based on our \revise{definition of misuse}, we mark the paper to be \bluen{``correct''} if all DR techniques used in the paper align with the tasks. We mark the paper to \redn{``fully misuse''} or \redt{``partially misuse''} DR when entire or part of DR techniques are used for unsuitable tasks, respectively (\revise{See Appendix D for examples}). 
\revise{We do so by carefully reviewing the justification for selecting each DR technique, the main task descriptions, and the target tasks described in user and case studies.}
Then, for each analytic task, we compute the proportion of papers that employ suitable DR techniques relative to all papers addressing the task.
As a result (\autoref{fig:error-tasks}), we find that tasks supported by local techniques have a high proportion of proper usage. However, tasks requiring global techniques have a substantially lower rate of proper usage.
This indicates that researchers properly employ local techniques (\tsne and \umap) when required, while also correctly avoiding global techniques for these tasks.
However, tasks requiring global techniques have a substantially lower rate of proper usage, indicating that \tsne and \umap are misused even for unsuitable tasks.

This tendency causes \tsne and \umap to dominate the misuse of DR techniques in practice. We find that \tsne is the most widely misused DR technique in our list of papers (\autoref{fig:misuse_summary}a). The misuse of \umap has recently increased rapidly, making it the runner-up (\autoref{fig:trend}b). We also find that \umap has the highest misuse-to-usage ratio among the four major DR techniques (\autoref{fig:misuse_summary}b). In summary, \tsne and \umap are misapplied more frequently than other techniques, underscoring the need for increased efforts to address these misuses.

In addition, we observe that a similar pattern persists when we exclude papers published before 2016 and 2019---the years when two influential papers guiding researchers on the use of \tsne and \umap are released \cite{distill2016how, coenen19fiar} (Appendix C).
This indicates that the misuse of \tsne and \umap persists despite existing efforts to inform practitioners how to use these techniques properly.




\paragraph{(Finding 3) \tsne and \umap are used without \revise{rationales} or with improper \revise{rationales}}
We find that more than 40\% of papers do not explicitly justify their choice of DR techniques (\autoref{fig:reasonings}; H3), and this trend persists for \tsne and \umap. 
These papers often discuss the general need for DR or describe the techniques’ characteristics rather than explaining why specific techniques are chosen.
This result implies that practitioners may perceive DR technique selection---including the case of \tsne and \umap---as requiring less critical evaluation, suggesting a lack of clear understanding of the appropriate way of using DR. Our subsequent interviews (\autoref{sec:interstudy}) further reinforce this observation in the context of \tsne and \umap usage.

We also identify that faithfulness is more widely used to justify the use of \tsne and \umap compared to PCA and MDS, but is referred to even when these techniques are misused (\autoref{fig:reasonings}).
This result thus supports our implications that researchers often lack proper understanding of DR; they may not know that \tsne and \umap are faithful in preserving local structure but not in global structure (\autoref{sec:suitability}).

\begin{table}[t]
    \centering
    
    \caption{\textit{The demographics of the participants in our interview study with practitioners.} VA and DR denote years of experience in visual analytics and DR, respectively.}
    \vspace{-1.2mm}
    \scalebox{0.84}{
    \begin{tabular}{rlllllll}
    \toprule
     & \textbf{Occupation}  & \textbf{Age} &\textbf{Gender} &\textbf{Domain} & \textbf{VA}  & \textbf{DR} \\
    \midrule
        P1 & Professor &  35& Male & Visual Analytics & 10  & 7  \\ 
        P2 & Undergraduate & 22 & Female & Visual Analytics & 2  & 1  \\ 
        P3 & Research Scientist & 30 & Male & Computer Vision & $\cdot$ & 6  \\ 
        P4 & Research Scientist & 30 & Male & Signal Processing & $\cdot$ & 4  \\ 
        P5 & Ph.D. Student & 28 & Female & Visual Analytics & 6 & 5 \\ 
        P6 & Ph.D. Student & 28 & Male & HCI & $\cdot$ & 4    \\ 
        P7 & Ph.D. Student & 29 & Male & Chemistry & $\cdot$ & 4 \\ 
        P8 & Ph.D. Student & 24& Female & Visual Analytics &3 & 2 \\ 
        P9 & Research Scientist & 30 & Male & NLP &  $\cdot$ & 7 \\ 
        P10 & MS Student &  22 & Male & Visual Analytics & 2  & 1 \\ 
        P11 & Staff Engineer & 34 & Male & Visual Analytics & 2  & 2 \\ 
        P12 & Postdoc. & 34 & Male & Bioinformatics & $\cdot$ & 4  \\ 
    \bottomrule
    \end{tabular}
    }
    \label{tab:demo}
\end{table}




\subsection{Takeaways}
\label{sec:littake}

The following are key takeaways from our literature review.

\paragraph{Misuse of \tsne and \umap exists}
We quantitatively verify that \tsne and \umap are widely misused in practice (Findings 2 and 3).

\paragraph{Researchers may have a limited understanding of how to use DR properly}
We find that researchers often use \tsne and \umap without proper justification or regard them as faithful for unsuitable tasks (Finding 3), implying limited awareness of various DR techniques and their suitability to analytic tasks. Indeed, the misuse of \tsne and \umap itself suggests that practitioners do not know how to properly use DR. 
\revise{The common misuse of PCA and MDS, affecting around 50\% (\autoref{fig:misuse_summary}), further indicates that this issue is not limited to \tsne and \umap, but reflects a broader lack of DR literacy. Still, the misuse of \tsne and \umap is particularly salient because of their popularity.}
Our interview study (\autoref{sec:interstudy}) reaffirms this observation.

\paragraph{Researchers may lack the motivation to use \tsne and \umap properly} 
We find that papers misusing \tsne and \umap have passed peer review and been published in major visualization conferences and journals.
This finding suggests that reviewers frequently overlook the importance of using DR appropriately. 
Furthermore, the frequent absence of clear justifications for selecting DR techniques (Finding 3) implies that researchers often do not recognize their importance.

\section{Interview Study with Practitioners}

\label{sec:interstudy}


We aim to investigate why practitioners misuse \tsne and \umap (\otwo).
To this end, we conduct interviews with practitioners who regularly use DR techniques for their research or visual analytics work.




\subsection{Study Design}

\label{sec:interstudydesign}

\noindent
\textbf{Participants and recruitment.}
We want to diversify our participants' experience with DR. We first aim to achieve diversity across the domains in which participants work.
To do so, we recruit both visual analytics researchers and domain researchers with experience visually analyzing and presenting their data using \tsne and \umap. 
For visual analytics researchers, we randomly select papers from our literature review (\autoref{sec:litreview}), prioritizing diversity in target data and problem domains. We then contact either the first or the corresponding author via email to increase diversity in participants' expertise and visualization literacy.
For domain researchers, we ensure they come from distinct disciplines.
We recruit participants from a local university through an internal web community with snowball sampling \cite{goodman61ams}.
We interview 12 participants (six visual analytics researchers and six domain researchers; \autoref{tab:demo}).


\paragraph{Interview protocol}
We interview participants in a semi-structured manner.
We first ask participants for consent. 
In doing so, we clarify that we aim to identify potential risks in participants' reasoning and do not intend to blame participants or other researchers.
We then ask participants to describe (1) their expertise in DR, \tsne, and \umap, (2) their experience and justifications in using \tsne and \umap, and (3) the difficulties that occur while using these techniques (our questionnaire is in Appendix B). 
The interviews are conducted via a recorded Zoom call. We compensate participants with approximately 15 USD. 
All interviews are completed within 40 minutes.




\subsection{Analysis Procedure}

\label{sec:interviewanalysis}

One coder initially analyzes the interview transcripts using thematic analysis \cite{joffe11qrm}. Each code focuses on describing practitioners’ behaviors when analyzing high-dimensional data using DR. The coder extracts participant quotes as initial codes and then hierarchically clusters them into two levels based on semantic similarity. Each lower-level cluster is grouped into a subtheme, which is subsequently abstracted into the final set of themes representing higher-level conceptual categories. 
Note that we group codes as subthemes only when supported by quotations from at least two different participants. 
Finally, two additional authors and the initial coder collaboratively review and refine the analysis.

\subsection{Findings}

\label{sec:addfindings}

We identify three themes and elaborate on each as a separate finding.

\paragraph{(Finding 1) Practitioners have limited literacy on DR}
We find that the participants have difficulty understanding not only \tsne and \umap but also other DR techniques, consistent with our takeaways from the literature review (\autoref{sec:littake}).
For instance, five participants report difficulties in choosing the final projection for deployment, as the outcomes vary significantly depending on hyperparameter configurations. Three of these participants admit that they are unsure of the proper way to set these hyperparameters. Second, participants often do not have sufficient knowledge of alternative DR techniques. For example, four participants say they are unaware of methods other than \tsne, UMAP, and PCA, and two of these participants mention that they can hardly tell the differences among these three techniques.

We also notice that practitioners' limited literacy makes them ``immune'' to using \tsne and \umap for their research. 
Participants mention that the popularity of \tsne and \umap makes them less likely to invite criticism of their analytic results or systems. 
For example, three participants mention using UMAP because they feel that alternative techniques may expose their paper to reviewer criticism. 
This finding aligns with our observation that many of the papers do not provide adequate justification for using \tsne and \umap (\autoref{sec:quantianal}).

\paragraph{(Finding 2) Practitioners receive misleading suggestions}
We find that practitioners regularly receive potentially misleading suggestions to use \tsne and \umap.
We identify three primary sources:

\paragraphit{Fellow researchers}
Five participants indicate that their fellow researchers recommend using \tsne and \umap. 
Two participants, in particular, mention that they unreservedly follow recommendations from their advisors or seniors. 
For example, a participant comments: \textit{``My advisor recommended UMAP, and I used it without verification''}.

\paragraphit{Research papers}
Two domain researchers state that they used \tsne and \umap after they frequently encountered them in research papers in their domains (bioinformatics and chemistry). 
One participant mentions that he regularly uses UMAP because it is frequently cited in recent publications in his domain (bioinformatics) for the same purpose.

\paragraphit{Language models}
Two participants mention that they ask large language models (LLMs) to recommend a DR technique to use, where the models suggest \tsne, \umap, and PCA. 
Both participants report using ChatGPT for this purpose. 

\vspace{4pt}
These suggestions can be misleading as they typically lack grounded evidence. For example, research papers often misuse \tsne and \umap (\autoref{sec:litreview}), so relying on these papers can lead to erroneous applications of these techniques. As LLMs are trained on massive text corpora, their frequent recommendations of \tsne and \umap can be interpreted as reinforcing practitioners' bias toward these techniques. 
Despite the existence of materials that inform the proper use of DR (\autoref{sec:weaknesses}), practitioners' reliance on such misleading suggestions indicates insufficient motivation to engage with these resources.

\paragraph{(Finding 3) Practitioners often cherry-pick hyperparameters}
We observe another misuse pattern: cherry-picking of hyperparameters. Eight participants report having experience manually tuning the hyperparameters of \tsne and \umap. 
They report that they aim to achieve either an interpretable or an aesthetically pleasing projection, with well-separated classes or clusters. 
Four participants report tuning hyperparameters without understanding their effects on projection results.


\begin{table}[t]
    \centering
    
    \caption{\textit{The demographics of the participants in our interview study with DR experts.} Exp. and Pub. denote the years of research experience and the number of publications related to DR, respectively.}
    \vspace{-1.2mm}
    \scalebox{0.81}{
    \begin{tabular}{rllllll}
    \toprule
     & \textbf{Occupation}  & \textbf{Age} & \textbf{Gender} & \textbf{Exp.} & \textbf{Pub.} & \textbf{Subdiscipline in DR} \\
    \midrule
        P1 & Assoc. Prof. & 34 & Male & 12 & $>$10 & Faithfulness, Scalability\\  
        P2 & Assoc. Prof. & 48 & Male & 20 & $>$10 & Algorithm, Evaluation\\ 
        P3  & Assist. Prof. & 39 & Male & 13 & $>$10 & Interpretation\\
        P4 & Assoc. Prof.  & 40 & Male & 20 & $>$10 & Interpretation\\ 
        P5 & Assist. Prof. & 36 & Male & 11 & 4 &  Practical use of DR \\
        P6 & Ph.D. Student & 26 & Male & 4 & 3 & Stability, Faithfulness  \\ 
        P7 & Ph.D. Student & 26 & Female & 3 & 3 &  Faithfulness \\ 
        P8  & Ph.D. Student & 28 & Male & 6 & 8 & Visual analytics for DR  \\ 
    \bottomrule
    \end{tabular}
    }
    \label{tab:demoexpert}
\end{table}

\subsection{Takeaways}

\label{sec:intertake}

We identify two key takeaways from the interview study:

\paragraph{Practitioners lack understanding of how to use DR properly}
The \revise{interview study shows that practitioners may have} a limited understanding of DR. Many participants (1) hold erroneous beliefs about the faithfulness of \tsne and \umap (Finding 1), (2) do not know how to select appropriate techniques (Finding 1), and (3) do not know how to properly set hyperparameters (Finding 3).

\paragraph{Practitioners lack motivation to use DR properly}
The interview study suggests that practitioners are unaware of how to use DR effectively and lack motivation to do so, aligning with our third takeaway from the literature review (\autoref{sec:littake}). For example, the perception of \tsne and \umap as ``immune to criticism'' arises from reviewers' insufficient interest in the proper use of DR. 
\section{Interview Study with DR Experts}

\label{sec:interviewexperts}

We want to understand why previous efforts are not effective in addressing the misuse (\othree) (\autoref{sec:checklist}). 
To this end, we interview visualization researchers whose primary research focus is DR. 

\subsection{Study Design}

\label{sec:expertdesign}

\noindent
\textbf{Participants and recruitment.}
We establish two objectives in recruiting the expert researchers.
First, we want our experts to have a sufficient \textbf{expertise} in (1) the underlying mechanism of DR and (2) how it is used in practice for visual analytics. 
Experts without the former may struggle to understand essential concepts for our problem, e.g., the rationale behind different DR techniques and their alignment with analytic tasks.
Conversely, experts without the latter may provide limited insights into addressing the misuse.
Second, we aim for our pool of experts to be sufficiently \textbf{diverse} in their experience with DR. 

To achieve these goals, we recruit experts by randomly sending emails to the authors of papers that address the misuse of DR, listed in our related work section (\autoref{sec:weaknesses}).
We contact one of the corresponding authors and the first author to further diversify expertise and research experience. \autoref{tab:demoexpert} depicts the demographics of our experts.

\paragraph{Interview protocol}
We conduct one-on-one interviews with each expert. One experimenter manages all experiments. 
After the experts sign the consent form, the experimenter verbally describes the purpose of the experiment and how it will proceed. 
Then, we share misuse patterns we find from the literature review (\autoref{sec:litreview}) and the interview with practitioners (\autoref{sec:interstudy}).
Subsequently, we ask interviewees to complete the following questionnaires, asking why the previous efforts in the literature to address the misuse have hardly achieved the goal.
\begin{itemize}[leftmargin=9pt]
    \item \textit{``Why do you think these misuses persist despite the large body of literature that informs practitioners not to do so?''}
    \item \textit{``Why do you think practitioners are not motivated to read existing materials?''}
\end{itemize}
Each expert answers each question verbally.
We wrap up the interview by requesting the experts to share their thoughts on how the misuses can be addressed. 
We do not limit the interview duration to fully elicit the experts' insights, but all interviews end within 40 minutes. We compensate experts with the equivalent of 20 USD for their participation.

\subsection{Analysis Procedure}

We apply the same thematic analysis procedure used in the previous interview study (\autoref{sec:interviewanalysis}), resulting in a set of themes and corresponding subthemes. 
The only methodological difference is that, given the smaller sample size of eight experts, we permit the formation of a subtheme even when it is supported by a single relevant code.

\subsection{Findings}

\label{sec:expertfinding}

All experts agree that misuse persists despite existing efforts and share the following thoughts on underlying causes. 

\paragraph{(Finding 1) Cultivating DR literacy is not easy}
Experts note that cultivating DR literacy is not easy, even for trained researchers. Two experts note that although survey papers exist to comprehensively inform the proper use of DR, these resources are difficult for novices to understand. 
One expert notes that understanding survey papers ultimately requires reading individual papers, which is a demanding task. 
This finding suggests that existing efforts to instruct practitioners---organizing and presenting information scattered in diverse papers \cite{jeon25chi} (\autoref{sec:weaknesses})---hardly reduce the inherent difficulty of learning DR.

\paragraph{(Finding 2) Libraries are promoting the misuse of \tsne and \umap}
Five experts note that highly polished and well-maintained libraries that serve \tsne and \umap intensify the misuse. They say that although practitioners want to test other DR techniques, they can hardly find and execute implementations, coming back to \tsne and \umap.
Of these, three experts note the need for new libraries that serve diverse DR techniques. 
Two experts also mention that not only DR techniques but also other infrastructures for utilizing DR, such as preprocessing or evaluation metrics, should be provided as libraries.


\paragraph{(Finding 3) Intrinsic bias of human perception is promoting the misuse of \tsne and \umap}
Two experts casually mention the possibility that previous attempts fail because there exists an intrinsic bias towards well-separated clusters. One of them notes that in such a case, it is natural for \tsne and \umap to be widely used because these techniques exaggerate cluster structure \cite{lee11pcs, jeon25tvcg}.
\revise{This is because these techniques are designed to place neighboring points close together and non-neighboring points farther apart than they are in the original space (\autoref{sec:suitability}). }
The finding aligns with the recent finding of Doh et al. \cite{doh25arxiv} that there exists a perceptual factor that makes practitioners prefer DR projections with clearly separated classes or clusters. 

\paragraph{(Finding 4) Mitigating the misuse is an urgent problem}
Three experts note that we should urgently address these misuses because they may compromise the reliability of scientific discoveries based on visual analytics. P5 states: \textit{``False positives (due to the misuse of DR) are undoubtedly occurring somewhere at this very moment.''} Two experts mention that the urgency intensifies because misuses can propagate through citation networks; they emphasize that intervention is needed before these misuses become de facto standards. Both experts note that \tsne and \umap appear to have already approached this status, warning that mitigation will become increasingly difficult over time.

\subsection{Takeaways}

\label{sec:experttake}

\paragraph{We should move beyond papers}
We identify the limitations of conventional efforts to mitigate the misuse of DR, which are largely based on academic papers (Finding 1). Moreover, experts emphasize the need for technical solutions to mitigate this bias (Finding 2). This suggests that we need to invest efforts that extend beyond traditional academic approaches to make proper use of DR as the norm.

\paragraph{\revise{We need advertisement and maintenance of tools}}
\revise{Though experts note the lack of libraries for alternative DR techniques and related infrastructures, many of these tools already exist. For example, a wide range of alternative DR techniques are available through libraries \cite{jeon25tvcg, wang21jmlr, amid22arxiv}, and libraries have been released for computing evaluation metrics \cite{jeon23vis, machado25visgap}. In addition, approaches to support DR optimization and reduce cherry-picking have been developed, with their implementations made publicly available \cite{jeon25arxiv, espadoto21tvcg}. The discrepancy between the existing tools and our interview findings suggests that these tools remain insufficiently adopted despite their availability. This, in turn, highlights the need for greater dissemination of such tools, as well as maintenance efforts that make them easier to discover, install, and use in practice. } 

\paragraph{We should act immediately}
The interview suggests that immediate action is necessary to address the misuse. Experts warn that the misuse may already have passed the point of no return (Finding 4).

\section{Discussion}

We discuss future directions to facilitate the proper use of DR.

\subsection{Human-Centered Automation}

\label{sec:checklist}

We claim that human-centered automation \cite{shneiderman22book} can be a promising direction to address the misuse of \tsne and \umap.
Our literature review and interviews reveal two underlying causes of misuse: (1) practitioners often lack the knowledge required to use DR appropriately, and (2) they have limited motivation to acquire that knowledge.
Addressing these issues and enabling sustained, proper use of DR can benefit from a human-centered use of automation \cite{shneiderman22book}: automating repetitive, redundant tasks (e.g., DR configuration) to make it easier and less burdensome, while preserving practitioners’ agency over final decisions (e.g., the selection of DR technique) so they can cultivate DR literacy through the configuration process.
We discuss the validity of this argument in a dialectical manner.

An intuitive solution to prevent the misuse of a method is to automate the parts that practitioners struggle to use correctly. For example, AutoML \cite{feurer15nips} has been widely studied to reduce the burden of machine learning practitioners by automatically tuning hyperparameters.
Likewise, it seems beneficial to automate key configuration choices for DR, such as selecting techniques and tuning hyperparameters, in ways that adapt to the analytic tasks and contexts.
\revise{One plausible direction is to leverage LLMs, given that practitioners regularly ask them to select DR techniques (Finding 2 of \autoref{sec:addfindings}). For example, equipping LLMs with a knowledge base on DR can enable more reliable recommendations while preserving the convenience of LLM-based assistance.} 

However, we should remain attentive to the risks of automation \cite{shin25tvcg, shneiderman22book}.
If a system makes end-to-end decisions, practitioners may be less motivated to learn to use DR properly, hindering the cultivation of their DR literacy. 
For example, if hyperparameter tuning is fully automated, practitioners may miss opportunities to learn the underlying mechanisms of DR techniques.
\revise{This problem may be amplified by the ease of use of automated solutions, for example, through simplified APIs, which allow users to obtain immediate benefits by applying techniques as-is and may further reduce their motivation to develop deeper DR literacy.
}
This has two disadvantages. 
First, the lack of DR literacy will limit the ability to critically scrutinize poorly configured DR projections, potentially undermining the credibility of scientific communication. 
Second, it may also hinder a comprehensive analysis of underlying data structures.
The goal of DR-based visual analytics is not to identify a single optimal projection, but to understand a dataset's distribution by comparing various projections while considering their algorithmic properties, appropriate tasks, and context.
From this perspective, limited DR literacy may limit practitioners' ability to conduct a thorough analysis even with fully automated DR. 

We still do not know the optimal way to use automation to both reduce the technical burden and preserve their agency.
We argue that a key direction to address this gap is to assess practitioners' DR literacy and provide adaptive support accordingly.
For example, we envision authoring assistants \cite{shin25tvcg, kim25pvis} that dynamically infer a user’s DR literacy during configuration and guide the process, e.g., by issuing contextual warnings when potentially improper settings are detected. 
A central challenge is to reliably estimate practitioners' literacy level. Decomposing DR literacy into measurable components and designing assessment methods will be critical steps toward addressing this challenge.

\subsection{Facilitating Discourse}

\label{sec:discussuse}

We anticipate a natural follow-up question to our warning against the misuse of \tsne and \umap: \textit{``Which techniques should we use instead?''}
Using PCA or MDS for global tasks and \tsne and \umap for local tasks can serve as a reasonable heuristic. 
However, this approach is not universally valid because each technique has its own limitations. For instance, the faithfulness of PCA degrades for non-Gaussian data distributions \cite{jolliffe05esbs, han12nips}. MDS suffers from quadratic time complexity.

We thus emphasize the importance of encouraging practitioners to cultivate their DR literacy, i.e., to understand the strengths and limitations of diverse DR techniques and to select them based on the given analytical tasks and contexts. For example, specialized techniques such as Densmap preserve density \cite{narayan21nature} (\autoref{fig:teaser}) while achieving competitive performance in preserving local structure, being suitable for scenarios in which tasks investigating global and local structure are both important \cite{jeon25tvcg}.
Moreover, other contextual factors play a role: when responsiveness is important (e.g., in real-time streaming data analysis \cite{shilpika22tvcg}), progressive DR techniques \cite{jo20tvcg} are preferable.

While efforts and studies to enhance DR literacy have been ongoing, our findings indicate that simply publishing guidance or survey papers is not sufficient (Finding 2 in \autoref{sec:quantianal}; Finding 1 and 2 in \autoref{sec:expertfinding}). 
  In response, we recommend moving beyond traditional text-based dissemination.
  Organizing workshops, panels, or tutorials at visualization, HCI, and machine learning conferences can help elevate the discourse and emphasize the importance of using DR techniques responsibly.
  Recent tutorials in EuroVis 2025 \cite{eurovis25evt} and CVPR 2025\footnote{\url{https://cvpr.thecvf.com/virtual/2025/tutorial/35919}} resonate with this path.
Investigating how these efforts synergize with human-centered automation (\autoref{sec:checklist}) will also be an intriguing future direction.

\subsection{Investigating Domains Outside of Visual Analytics}
In this research, we review relevant literature in the field of visual analytics. 
By doing so, we reveal that visual analytics researchers often use \tsne and \umap inappropriately. 
\revise{This finding is surprising, as the visual analytics community is arguably expected to have an in-depth understanding of how to use DR appropriately. It further suggests that misuse may be even more prevalent outside the visual analytics community.}
Our interview with domain researchers \revise{partially investigates these fields} (\autoref{sec:interstudy}), but they cover only four areas: machine learning, HCI, chemistry, and bioinformatics.
Examining the use of DR in domains beyond visualization could help us reach a wider audience and uncover more broadly applicable solutions. Cashman et al. \cite{cashman25arxiv} recently present a seminal step in this direction by executing a literature review on four domains outside visualization (Physics, Chemistry, Biology, and Business).
However, their investigation is based on a literature review and therefore hardly uncovers the in-depth rationales or challenges underlying their findings, unlike our combined approach of literature review and interview study.
Conducting additional interviews with domain researchers to probe the underlying causes of the challenges reported in their study will be a promising avenue for future research.


\subsection{Extending to Other Machine Learning Techniques}

We provide a detailed understanding of the phenomenon in which practitioners frequently misuse \tsne and \umap, often treating these techniques as one-size-fits-all solutions. We find that even experts in data visualization are not immune to these issues (\autoref{sec:interstudy}).

We claim to investigate whether a similar tendency exists in the application of other machine learning and AI techniques. For instance, although we cannot always guarantee the faithfulness of LLMs (e.g., due to hallucinations \cite{zhang23arxiv}), they are commonly used in many applications without proper performance evaluation, often justified by their perceived ability to generate human-like responses \cite{he25tvcg}. Investigating whether LLMs are applied to appropriate tasks and properly parameterized is thus necessary. 
 As our research does for \tsne and \umap, such examinations will help develop solutions that support the proper use of machine learning for visual analytics and stimulate related discourse.

 \subsection{\revise{Immediate Actions}}

\revise{
Although we identify future research directions that could help mitigate the misuse of DR, pursuing them will require substantial time and effort. To complement, we provide a simple workflow guideline that can help practitioners immediately avoid misusing DR in visual analytics.}

\paragraph{\revise{Step 1: Understand your tasks}}
\revise{Practitioners should understand their target task. 
They can review the analytic tasks we list in \autoref{sec:analtasks} and \autoref{fig:tasks}. 
They can then identify the tasks that align with their objectives.}

\paragraph{\revise{Step 2: Find DR techniques and metrics that match your task}}
\revise{
Practitioners are required to find DR techniques and metrics that match their tasks. 
In terms of techniques, practitioners can naively rely on the alignment between DR techniques and tasks that we provided in \autoref{sec:suitability}. 
Ideally, practitioners can find matching techniques by reviewing papers that propose or benchmark DR techniques, comparing their pros and cons \autoref{tab:tasks}.
For metrics, practitioners can rely on the literature review and taxonomy on evaluation metrics \cite{jeon23vis, machado25visgap}.
}

\paragraph{\revise{Step 3: Optimize hyperparameters}}
\revise{
In this step, practitioners should iteratively execute the selected DR techniques to optimize their hyperparameters. 
This is achieved by testing various hyperparameter settings and by evaluating the resulting projections using the selected DR evaluation metric.
Recent studies \cite{jeon25arxiv, espadoto21tvcg} have released code for optimizing hyperparameters efficiently, while traditional approaches like random search and Bayesian optimization \cite{snoek12nips} can also be used. 
}

\paragraph{\revise{Step 4: Execute post-hoc investigation}}
\revise{
Even when projections are properly configured and optimized, they cannot fully avoid distortion \cite{nonato19tvcg, jeon21tvcg}. Therefore, it is important to understand where and how such distortion occurs \cite{lespinats11cgf, eurovis25evt}. We recommend using distortion visualizations that mark poorly projected points \cite{lespinats11cgf, jeon21tvcg}. Such visualizations are already available in existing libraries \cite{jeon23vis}.
}
\section{Conclusion}

We critically examine the misuse of \tsne and \umap in visual analytics.
Through the literature review and interview studies,
we verify the existence of the misuse, reveal why such misuse occurs. 
and identify that existing efforts fail to motivate practitioners to develop their DR literacy. 
Based on these findings, we suggest leveraging human-centered automation to mitigate DR misuse.
We also suggest future directions to address the misuse.

\revise{Our study highlights the need for broader investigations into the misuse of DR. Although the limited sample size of our studies may raise concerns about the generalizability, such small-scale qualitative studies have already revealed a range of important problems in how DR is used. We thus call for large-scale interview studies and systematic investigations in visual analytics and beyond the field as future work.}

\acknowledgments{%
This work was supported by the National Research Foundation of
Korea (NRF) grant funded by the Korean government (MSIT) (No.
NRF-2023R1A2C2005209), the Institute of Information \& communications Technology Planning \& Evaluation (IITP) grant funded by the Korean government (MSIT) [NO.RS-2021-II211343, Artificial Intelligence Graduate School Program (Seoul National University)], and by the SNU-Global Excellence Research Center establishment project. The ICT at
Seoul National University provided research facilities for this study.
Hyeon is in part supported by the Google Ph.D. Fellowship. 
}

\bibliographystyle{abbrv-doi-hyperref}

\bibliography{ref}








\end{document}